# Physics and Detectors at the LHC and the SLHC

Wesley H. Smith

*Physics Department, University of Wisconsin, Madison, WI, 53706, USA*

The capabilities of the ATLAS and CMS detectors being prepared for the LHC are reviewed. Examples of physics signals accessible during early running and during mature high luminosity LHC operation are examined. The planning and options for the LHC and these detectors to increase the luminosity to $10^{35}$ cm$^{-2}$s$^{-1}$ is presented. This upgrade, entitled the Super LHC (SLHC), would occur in the next decade. The resulting physics scope is discussed.

## 1. LHC, ATLAS AND CMS

The LHC is scheduled to start operation in 2007 with 43 colliding counter-rotating pairs of 7 TeV proton bunches increasing to 156, yielding a luminosity from $3 \times 10^{28}$ to $2 \times 10^{31}$. After a shutdown, another run of more than one year will consist of 936 bunches (75 ns spacing) with a luminosity of $1\text{-}4 \times 10^{32}$ transitioning to 2808 bunches (25 ns spacing) with a luminosity between $7\text{-}20 \times 10^{32}$. A long shutdown will precede running up to the full luminosity of $10^{34}$.

The ATLAS detector employs silicon pixels and strips and a straw-tube based transition radiation detector inside a superconducting solenoid with a 2 T field. Outside of this, a Pb-LAr electromagnetic calorimeter is enclosed by iron-scintillator (barrel) and Cu/W-LAr (forward) hadronic calorimetry. The ATLAS muon system is composed of muon drift tubes, thin gap chambers and resistive plate chambers embedded in a large array of 8 air-core toroid magnets.

The CMS detector uses silicon pixels and microstrips, PbWO4 crystal electromagnetic and brass-plastic scintillator hadronic calorimetry all inside a large 4T superconducting solenoid. The muon system is composed of drift tubes, cathode strip chambers and resistive plate chambers inserted between iron layers of the flux return.

Both ATLAS and CMS Level-1 (L1) trigger systems reduce the input crossing rate of 40 MHz to less than 100 kHz with custom processing of calorimeter and muon detector information. Further processing that eventually involves all detector data is used by their Data Acquisition (DAQ) systems to select an output rate of ~100 Hz of data archived.

## 2. LHC PHYSICS

### 2.1. LHC Startup

Initially with 10 fb$^{-1}$ of luminosity at startup there are several physics channels with the potential for discovery. A resonance that decays to leptons will generally provide a clear signature with a low background. New physics producing resonances could include a new Z', *e.g.* any new heavy gauge boson such as found in GUTs, dynamical EWSB or little Higgs models, or could include models with compact extra dimensions such as the Randall Sundrum model and other theories providing massive Kaluza-Klein excitations, *e.g.* Gravitons. These two types of new physics signals can be distinguished by examination of their angular distributions, where their spins set them apart.

Spectacular multi-jet, multi-lepton and missing energy signatures would also be produced by SUSY, with production of ~100 events per day at luminosities of $10^{33}$ for squark and gluino masses of ~ 1 TeV. However, there are uncertainties about the backgrounds. While the background calculations are improving, there will be a need to use data control samples matched to MC calculations to properly determine them and to take into account detector effects.

At LHC startup, for the low mass Higgs search, H → γγ and H → ZZ* → 4$l$ are the only channels with a mass peak having good (~1%) resolution. At higher mass, between 140 and 180 GeV, H → WW* → 2$l$2ν yields a high rate but no mass peak. This requires a good understanding of the SM background. In addition, qqH with H → WW* and ττ final states should be significant. Almost all of the allowed Higgs mass range is explored by ATLAS and CMS with 10 fb$^{-1}$, and with 30 fb$^{-1}$ is covered to more than 7σ over the whole range.

## 2.2. Mature LHC Program and its limits

When the LHC is running at design luminosity, if the Higgs has been observed then as additional luminosity accumulates, parameters such as the mass and couplings will need to be measured. This may require up to 300 fb$^{-1}$ or more. Due to the small branching ratio for clean final states, there are insufficient statistics to measure the Higgs self-coupling at the LHC. If SUSY is observed at the LHC then the masses and model will need to be determined, along with the connection to cosmology (*e.g.* dark matter), the impact on Higgs phenomenology and the SUSY breaking mechanism. Much of this is difficult with the LHC. If neither the Higgs nor SUSY is found, then other possibilities need investigation, such as strong $W_L W_L$ scattering, other EWSB mechanisms, extra dimensions, little Higgs models and Technicolor. Detecting and untangling such new physics may also be beyond the capabilities of the LHC.

## 3. LHC UPGRADE: SLHC

Assuming the nominal LHC luminosity profile, the LHC Interaction Region (IR) quadrupole life expectancy is less than 10 years and the statistical error halving time will exceed 5 years by 2012. Even if this profile is delayed by a year or two, this does not affect consideration of an LHC luminosity upgrade based on new low-β IR magnets for the middle of the next decade. In its present design, the LHC luminosity is limited by the beam dumping system, the machine collimation and protection systems, and electron cloud effects, which may constrain the minimum bunch spacing.

Upgrades to the LHC can be classified in 3 phases. Phase 0 would push the machine to maximum performance without hardware changes. This would involve colliding the LHC beams at points 1 and 5 (ATLAS and CMS) only with alternating horizontal and vertical crossings, and increasing the number of protons per bunch up to the beam-beam limit, yielding a luminosity of 2.3 x 10$^{34}$. The LHC dipole field could also be increased to its ultimate value of 9 T, raising the energy from 7 to 7.54 TeV. Phase 1 would be to modify the insertion quadrupoles and their layout to get a β* of 0.25 m, increase the crossing angle by ~1.4, increase the number of protons per bunch to its ultimate value, halve the longitudinal bunch width and double the number of bunches. This could conceivably yield a luminosity of 9.2 x 10$^{34}$ if electron cloud effects can be dealt with. Phase 2 would be an energy and luminosity upgrade involving injection system modifications, rebuilding the SPS with superconducting magnets and upgrading the transfer lines to inject into the LHC at 1 TeV and installing new 15 T dipoles, producing an energy of 12.5 TeV. Since the timescale for Phase 2 is the end of the next decade, this paper restricts the upgrade discussion to Phase 1, called the Super LHC (SLHC).

## 4. SLHC PHYSICS

The integrated luminosity of the SLHC program is roughly ten times that of the LHC or about 3000 fb$^{-1}$ per experiment. Such high statistics could be used to combine different Higgs production and decay modes to form ratios of the Higgs couplings to bosons and fermions. At the SLHC these ratios should be measurable with ~10% precision. The SLHC would provide an extension of the discovery domain for massive MSSM Higgs. The heavy Higgs observable region should increase by ~ 100 GeV. The range for SUSY discovery is also extended from 2.5 TeV to 3 TeV by use of high $E_T$ jets and missing $E_T$, which are not significantly degraded by the pile-up at the SLHC due to high thresholds. There is improved coverage of A/H decays to neutralinos and 4 isolated leptons, depending on the model or MSSM parameters assumed. The discovery mass range for new gauge bosons (*e.g.* sequential Z' model) extends from 5.3 TeV to 6.5 TeV. The mass reach for extra dimensions is also increased, as in the case of the Randall Sundrum model where it extends 30% or in the case of Kaluza Klein excitations of the γ or Z the direct observation limit from ATLAS and CMS combined increases from 6 to 7.7 TeV and observation through interference effects extends to 20 TeV.

## 5. SLHC DETECTORS

The SLHC presents a number of experimental challenges such as radiation damage to detectors, an increase in pile-up of additional overlapping events and a possible increase in the crossing frequency from 40 MHz to 80 MHz, which would partially offset this pile-up. Both the ATLAS and CMS tracking systems would need replacement. For the CMS calorimeter, the forward quartz fiber detector is sufficiently fast, but might require finer-grained information to provide a smaller trigger tower size. The CMS HCAL and ECAL have sufficient time and spatial resolution for 80 MHz operations, using their present 40 MHz sampling without significant modification. However, replacement of the high η calorimetry may be needed due to radiation damage. The ATLAS LAr calorimeter will experience more than a factor of 3 increase in pileup at the SLHC luminosity, which may require a change in the electronics shaping time to optimize noise performance. Some changes might be necessary for $|η| > 2$ to mitigate space charge effects. The ATLAS Tilecal will need additional study of calibration and energy corrections, as it may be challenging to extract a minimum ionizing signal amid the pileup background. It will suffer some radiation damage at high η, which may require partial replacement. The ATLAS and CMS muon systems both use RPCs that may not function at the SLHC luminosity, particularly at high η. The existing ATLAS Muon Cathode Strip Chambers (CSC) and Thin Gap Chambers will probably continue to be usable for triggering with some improvements and higher thresholds. The same is true for the CMS Muon Drift Tubes and CSCs.

The ATLAS and CMS trigger and DAQ systems would need significant modifications to operate at the SLHC. Due to the increased occupancy of each crossing at the SLHC, L1 trigger systems would experience degraded performance of the algorithms used for the LHC. The DAQ system would experience larger event sizes due to greater occupancy. If new, higher channel-count trackers replace the existing ones, then the increase would be greater. This would reduce the maximum L1 trigger rate for a fixed readout bandwidth. Additional trigger information from the rebuilt tracking systems could reduce the L1 trigger rate or could be used earlier in the higher level triggers. A new DAQ architecture exploiting new developments in commodity networking would help to address the SLHC challenge.

## Acknowledgments

The author wishes to thank S. Dasu, D. Denegri, A. De Roeck, G. Hall, B. Mellado, A. Nikitenko, F. Ruggiero, M. Spiropulu, and F. Zimmerman for advice and for providing information used in this paper.

Work supported by Department of Energy contract DE-FG02-95ER40896.